\documentclass{cjpsuppl}
\usepackage{graphicx}  
\newcommand{\pt}{$p_{T}$}
\newcommand{\ptmaxh}{$p_{T}^{\rm max}/2$}
\newcommand{\kt}{$K_{T}$}
\newcommand{\Et}{$E_{T}$}
\newcommand{\Dphi}{$\Delta \phi\,{}_{\rm dijet}$}

\newcommand{\GeVc}{GeV/c}

\newcommand{\TeVcc}{TeV/c$^2$}
\newcommand{\invpb}{pb$^{-1}$}
\begin{document}
\title{QCD at the Tevatron}
\authori{Marek Zieli\'nski}
\addressi{Department of Physics and Astronomy, University of Rochester, Rochester NY, USA}
\authorii{}    \addressii{}
\authoriii{}   \addressiii{}
\authoriv{}    \addressiv{}
\authorv{}     \addressv{}
\authorvi{}    \addressvi{}
\headtitle{QCD at the Tevatron}
\headauthor{Marek Zieli\'nski}
\lastevenhead{Marek Zieli\'nski: QCD at the Tevatron }
\pacs{}
\keywords{}
\refnum{}
\daterec{15 September 2004
}
\suppl{A}  \year{2004} \setcounter{page}{1}

\maketitle

\begin{abstract}
Recent measurements of selected QCD processes  at the Tevatron
are reviewed and confronted with theoretical calculations.
Results on inclusive jet production at large transverse momentum (\pt) are
compared to predictions from next-to-leading order (NLO) perturbative QCD (pQCD).
Kinematic distributions of jets with light and heavy flavor produced 
in association with
electroweak bosons are compared to expectations from
leading-order (LO) QCD calculations supplemented with
parton-shower models.
Properties of QCD radiation in hard-scatter events are
investigated using azimuthal correlations between two leading
jets in multi-jet events, while aspects of softer radiation
are examined through properties of energy flow within jets.
Finally, characteristics of soft interactions underlying
the hard scatter are explored in the context of tuning the parameters
of phenomenological models employed in QCD Monte Carlo
event generators.
\end{abstract}


The ongoing physics program at the Tevatron offers an unprecedented
opportunity to explore a wide variety of QCD issues at a level
of high precision.
In Run II, the collision energy increased
to $\sqrt{s}=1.96$ TeV, from 1.8 TeV in Run I, providing an
increase by a factor of 5 in the inclusive jet
cross section at \pt$\approx 600$ \GeVc. Now, in the
early phase of Run II, the CDF and D\O\ experiments 
have already collected
substantially more luminosity than in all of Run I.
As a consequence, the current reach in \pt\ 
of the inclusive
jet spectrum exceeds that of Run I by $\approx 150$ \GeVc,
and events with dijet masses in the range of 1.2--1.3 \TeVcc\
have been collected by both collaborations, probing
interaction length scales of 10$^{-19}$ m. With steadily
increasing data samples, we are all looking forward to
improved precision measurements and to exploiting
the discovery potential for new physics.

Production of jets at large \pt\ has been long recognized as
one of the most interesting probes of pQCD.
Within the pQCD framework, hard-scattering processes are described by
a convolution of partonic cross sections with parton distribution
functions (PDFs). Thus, measurements of high-$p_T$ 
interactions probe PDFs and the strong coupling constant, $\alpha_s$,
and test current phenomenological tools (among others, high-order matrix-element
calculations, models of parton showers, the interplay
between these two approaches, resummation techniques etc.).
Deviations from pQCD, not accommodated by uncertainties in these components,
may indicate presence of physics beyond the Standard Model 
(e.g., parton compositeness).

Production of objects at low and moderate $p_T$ is sensitive
to fragmentation, multiple-parton interactions, and to the 
nature of remnants of
the colliding particles not participating in the hard collision.
Studies of these effects
help develop phenomenological descriptions of such soft
physics and in the tuning of event generators.

The hard-scattered quarks and gluons in the final state fragment
into jets of particles that subsequently interact in the detectors.
Experimentally,
jets are usually defined by their energy deposition in multiple
calorimeter towers. For Run II, both CDF and D\O\ collaborations adopted
a new iterative
fixed-radius cone algorithm.
While Run-I jets consisted of $E_T$-weighted sums
of calorimeter towers,
the Run~II cone algorithm defines jets by summing 4-momenta of 
towers within a cone,
using midpoints between such jets as additional
seeds for the algorithm to minimize
sensitivity to infrared divergence of gluon radiation \cite{algos}.
Analogous procedure can be applied at parton or particle levels.
(CDF is also using their Run I version of the cone
algorithm JetClu \cite{jetclu} for easier comparison with previous results.)
Both collaborations continue using the \kt\ clustering algoritm,
based on relative transverse momenta between clustered objects.

\begin{figure}
\centerline{
\includegraphics[scale=0.35]{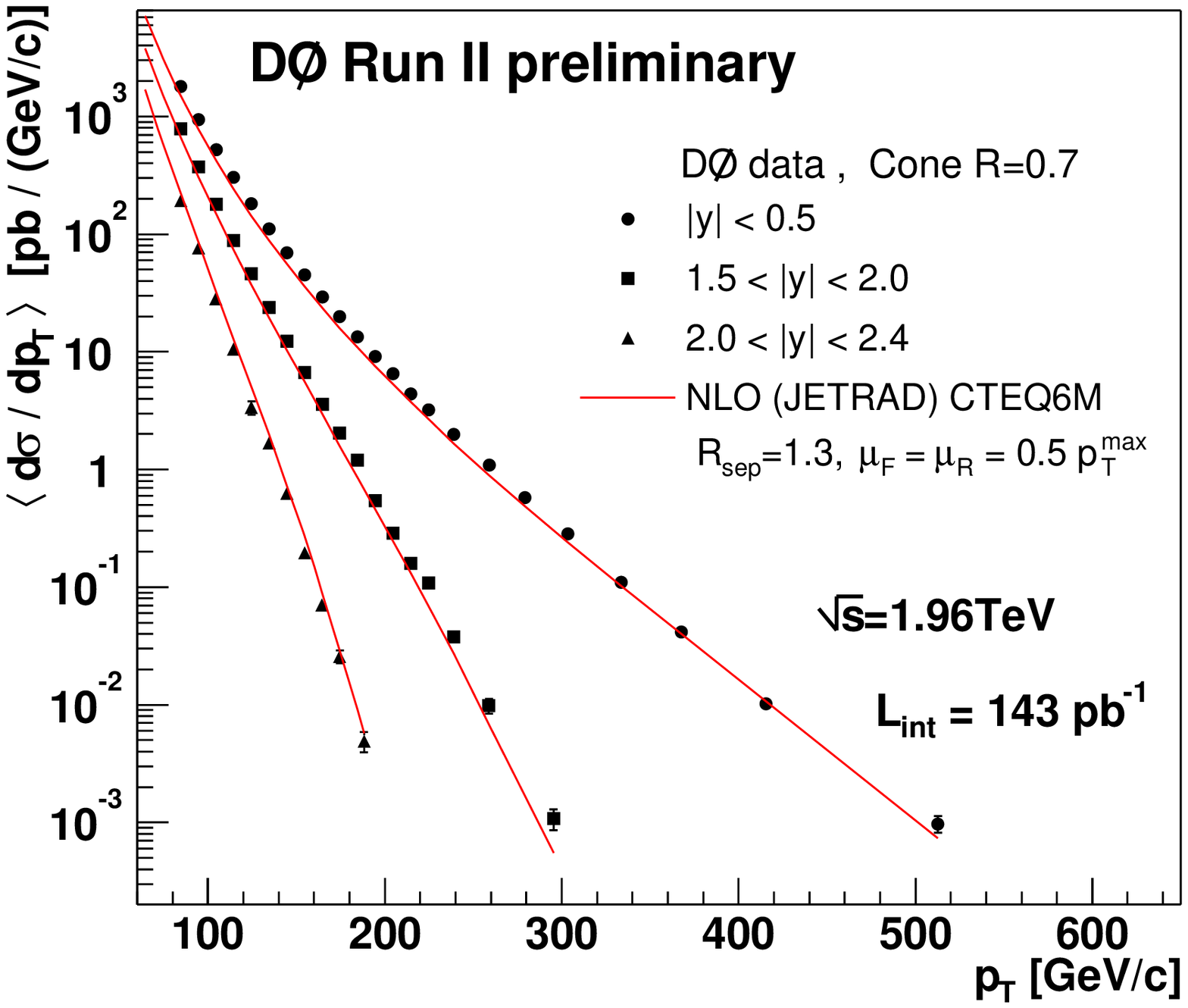}
\hskip -0.2in
\includegraphics[scale=0.35]{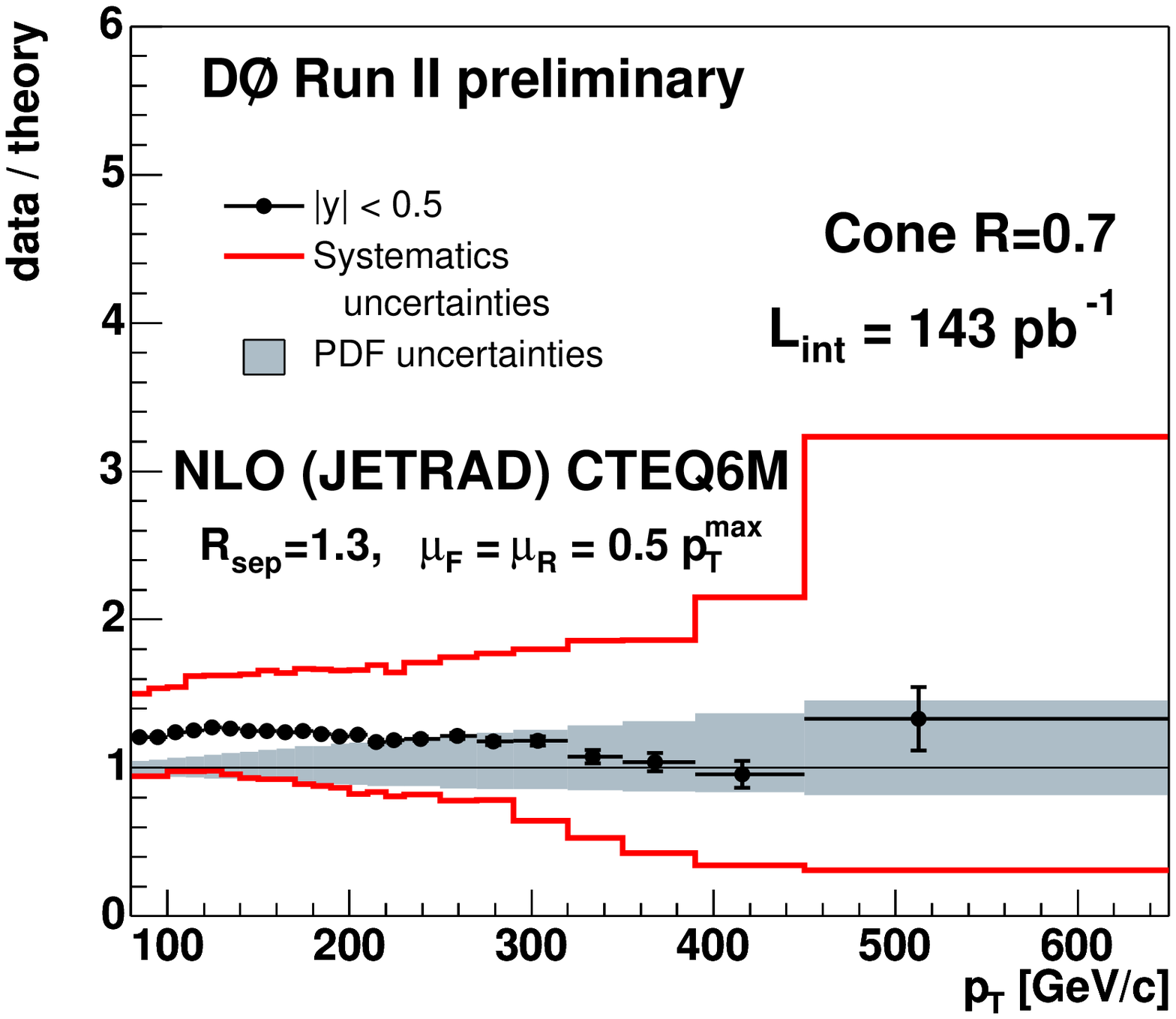}
}
\vskip -0.3in
\caption{\label{fig:data1} Left: Inclusive jet cross section measured by
D\O\ in three rapidity bins using the midpoint algorithm vs. jet \pt. 
The error bars are statistical only. 
Right: Ratio of the cross section for the central rapidity bin to
NLO pQCD; the outside band represents the current experimental
systematics, while the inner grey band shows uncertainty from PDFs.}
\end{figure}

\begin{figure}
\hskip -0.05in
\includegraphics[scale=0.35]{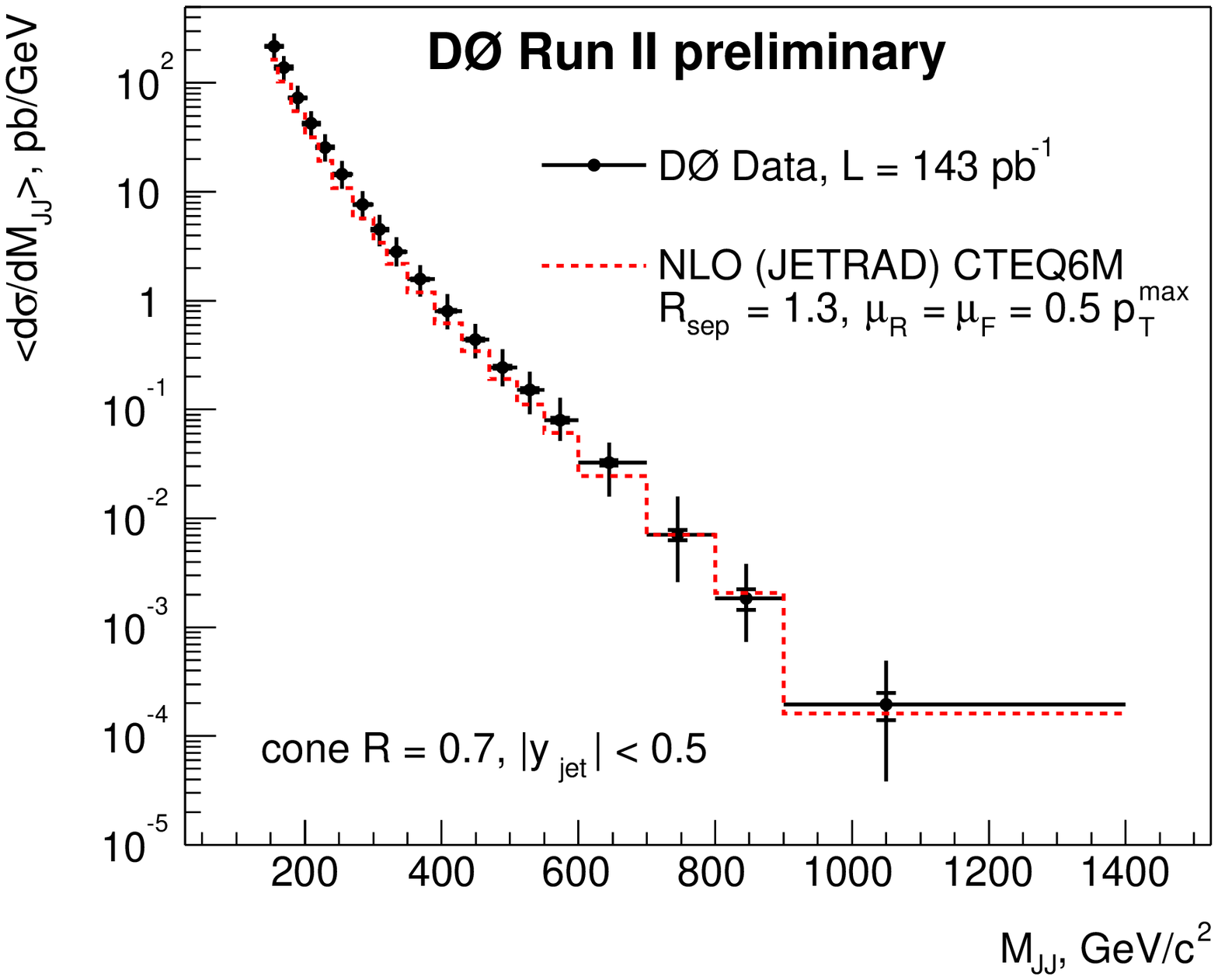}
\hskip -0.3in
\hfill
\includegraphics[scale=0.35]{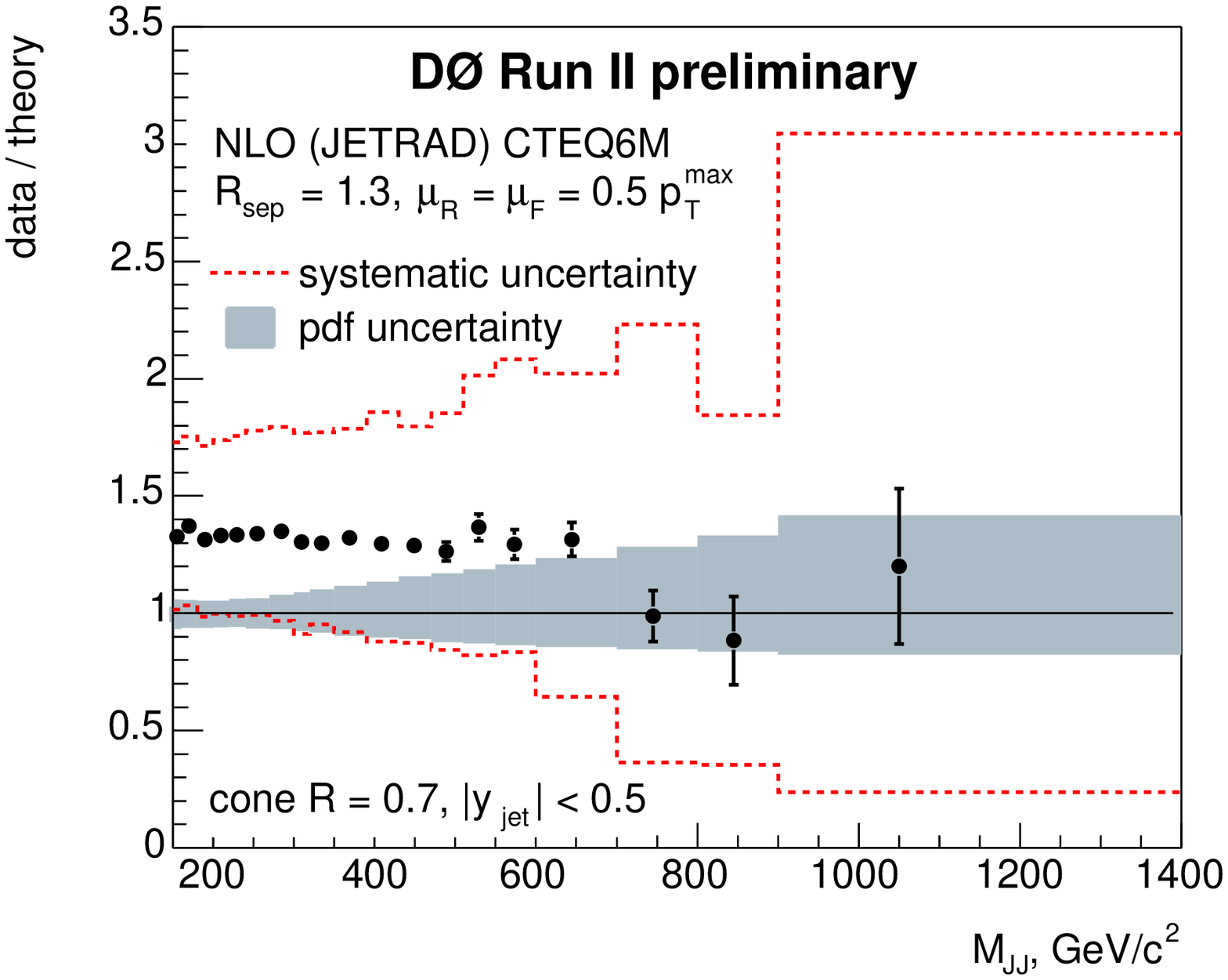}
\vskip -0.3in
\caption{\label{fig:data2} Left: Inclusive dijet cross section 
as function of dijet mass measured by D\O.
Right: Ratio to NLO theory; the bands show the experimental and 
PDF uncertainties.}
\end{figure}

The high-$x$ behavior of PDFs has been
scrutinized intensely for the past several years.
The Run I measurement from D\O\ \cite{levan} of
the inclusive jet cross section in several regions of pseudorapidity 
has provided powerful new information
for the global determinations of PDFs
by the CTEQ \cite{cteq6} and MRST \cite{mrst2002} groups.
Similar measurements are being pursued in Run II.
D\O\ has recently presented the preliminary inclusive jet cross section
as function of \pt\ in 3 rapidity ($y$) bins
(Fig.~\ref{fig:data1}), and a dijet cross section for
central rapidities $|y|<0.5$) (Fig.~\ref{fig:data2}), based on
data corresponding to an integrated luminosity of
143 \invpb. Given the large systematic errors,
the data are in good agreement with NLO pQCD \cite{jetrad},
for the parameters of the calculations 
noted in the figures.
So far, there is no indication of 
parton substructure or
presence of narrow resonances
decaying to jets. 
The uncertainties, shown as bands on the
plots of data/theory ratio, are dominated by the current
understanding of the jet energy
scale, and are expected to be
reduced in near future.

\begin{figure}
\centerline{
\includegraphics[scale=0.32]{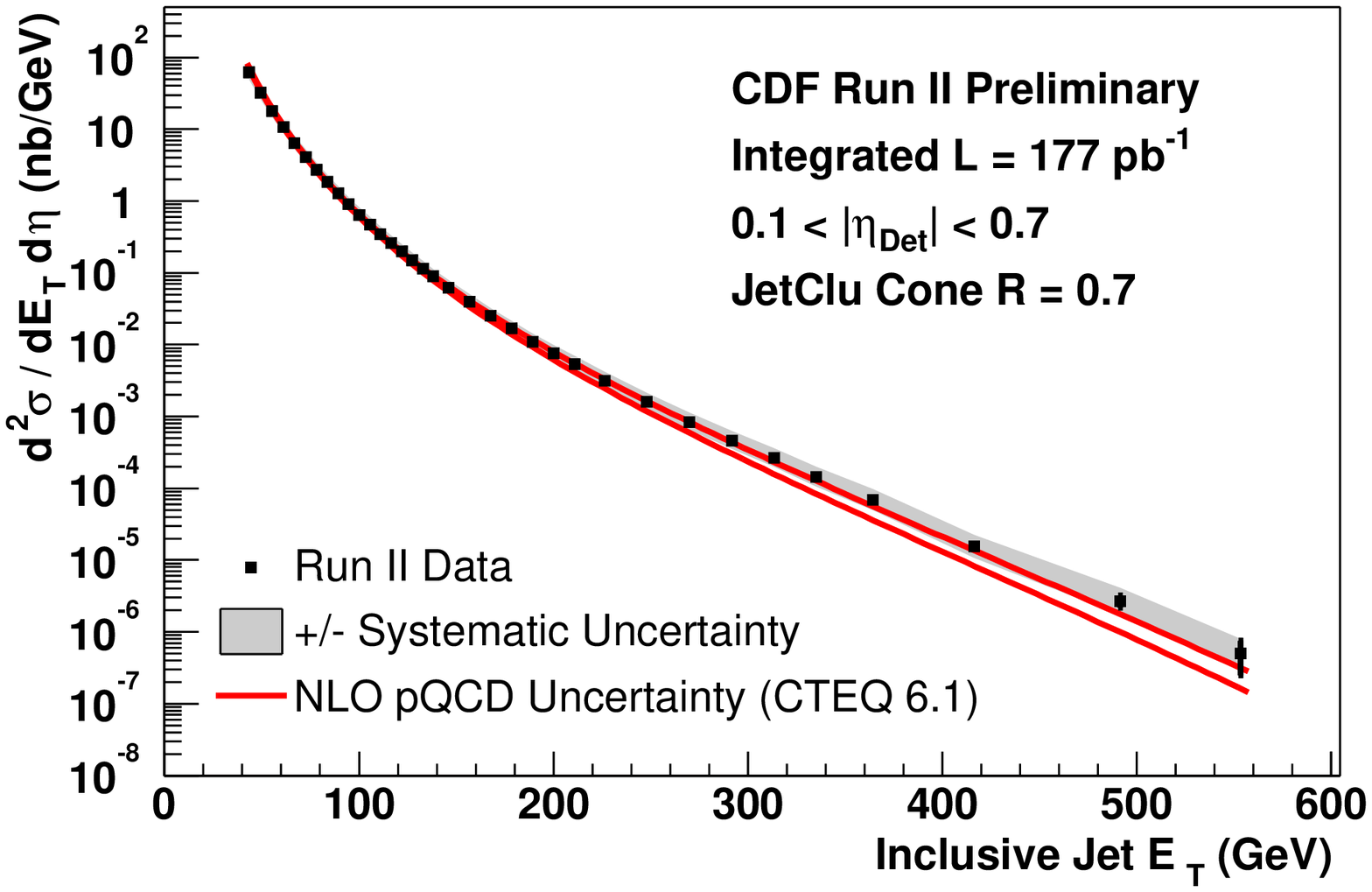}
\includegraphics[scale=0.32]{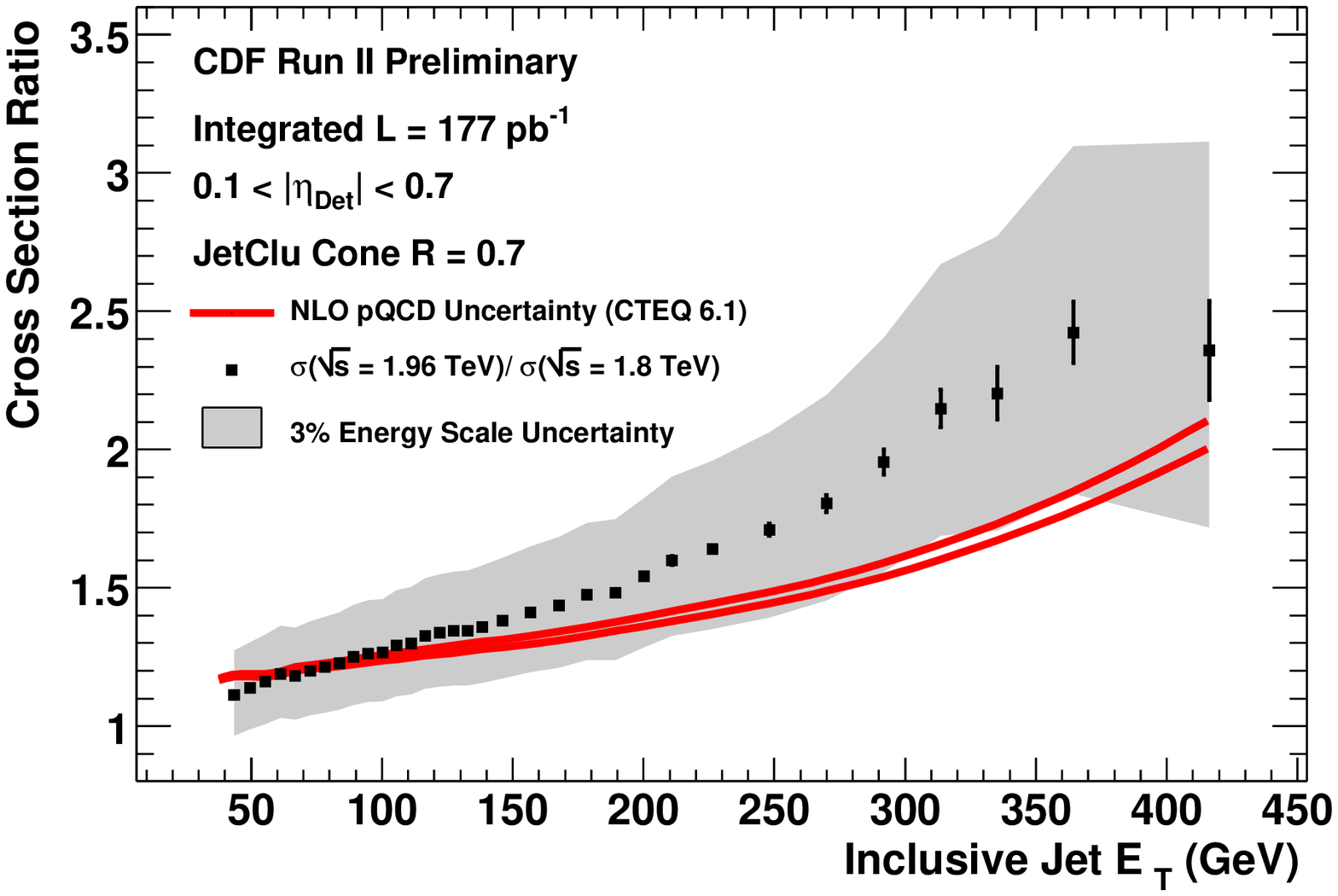}
}
\vskip -0.3in
\caption{\label{fig:data3} 
Left: Central inclusive jet cross section 
measured by CDF using the JetClu algoritm 
vs. jet $E_T$. Experimental and  PDF uncertainties
are also shown.
Right: Ratio of jet cross sections for Run II  to
Run I, compared to NLO pQCD.}
\end{figure}

CDF has presented a measurement of the inclusive
jet cross section for central rapidities ($0.1<|y|<0.7$)
using a data sample of 177 \invpb. It is compared
to NLO pQCD \cite{eks} and to Run I result
in Fig.~\ref{fig:data3} (both Run~II and
Run~I results were obtained with their JetClu algorithm).
The expected increase in the Run-II cross section as function 
of \pt\ for the larger $\sqrt{s}$
is clearly visible.
While there is reasonable agreement between data
and theory, it will be interesting to re-examine these
comparisons when experimental uncertainties become reduced.

\begin{figure}
\centerline{
\includegraphics[scale=1.05]{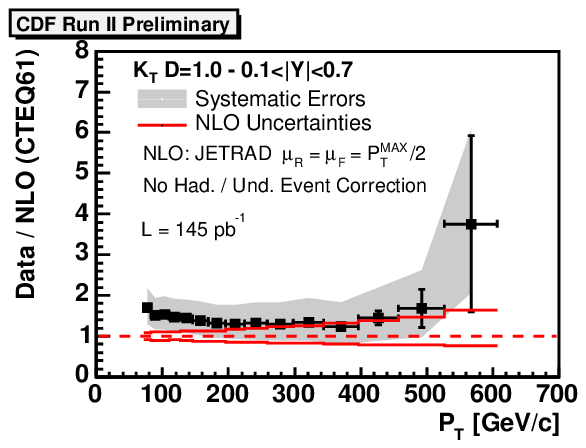}
\includegraphics[scale=1.05]{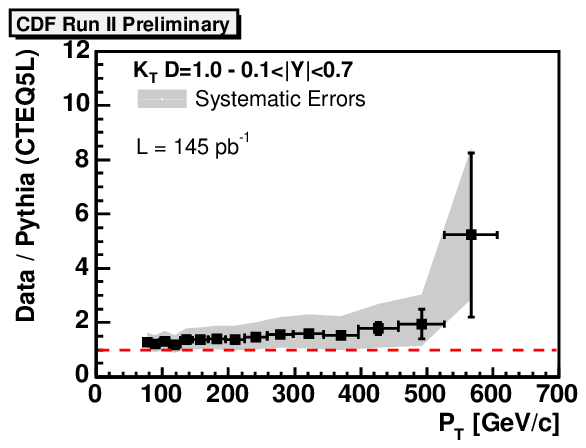}
}
\vskip -0.2in
\caption{\label{fig:data4} 
Left: Ratio of the central inclusive jet cross section 
measured by CDF using the \kt\ algoritm (with $D=1.0$) to  
NLO pQCD, as function of jet \pt.
Right: Similar ratio to prediction from {\sc pythia} (Tune A).}
\end{figure}

For a somewhat smaller data sample of 145 \invpb, CDF measured the
inclusive jet cross section over central rapidities
using the \kt\ clustering algorithm \cite{kt}
and jet-size parameter
values of $D=0.5, 0.7, 1.0$. The results
are presented in Fig.~\ref{fig:data4} for
$D=1.0$, in terms of 
ratios to NLO pQCD \cite{jetrad}, using  QCD scale \ptmaxh,
and to the prediction from {\sc pythia} \cite{pythia}. While the ratio 
to NLO exhibits an excess at lower values of \pt,
that grows as the value of $D$ increases (not shown),
no such effect is seen when comparing to {\sc pythia} with
parameters tuned to CDF data on the properties of the
``underlying event'' (so called Tune A, see end of paper).
This behavior may be due to the influence  of
particles in the underlying event on jet properties;
the measured cross section is not corrected for this contribution. 
The underlying-event effects are not present in the NLO calculation,
while they are included in the phenomenogical model implemented
in {\sc pythia}.

\begin{figure}
{
\includegraphics[scale=0.38]{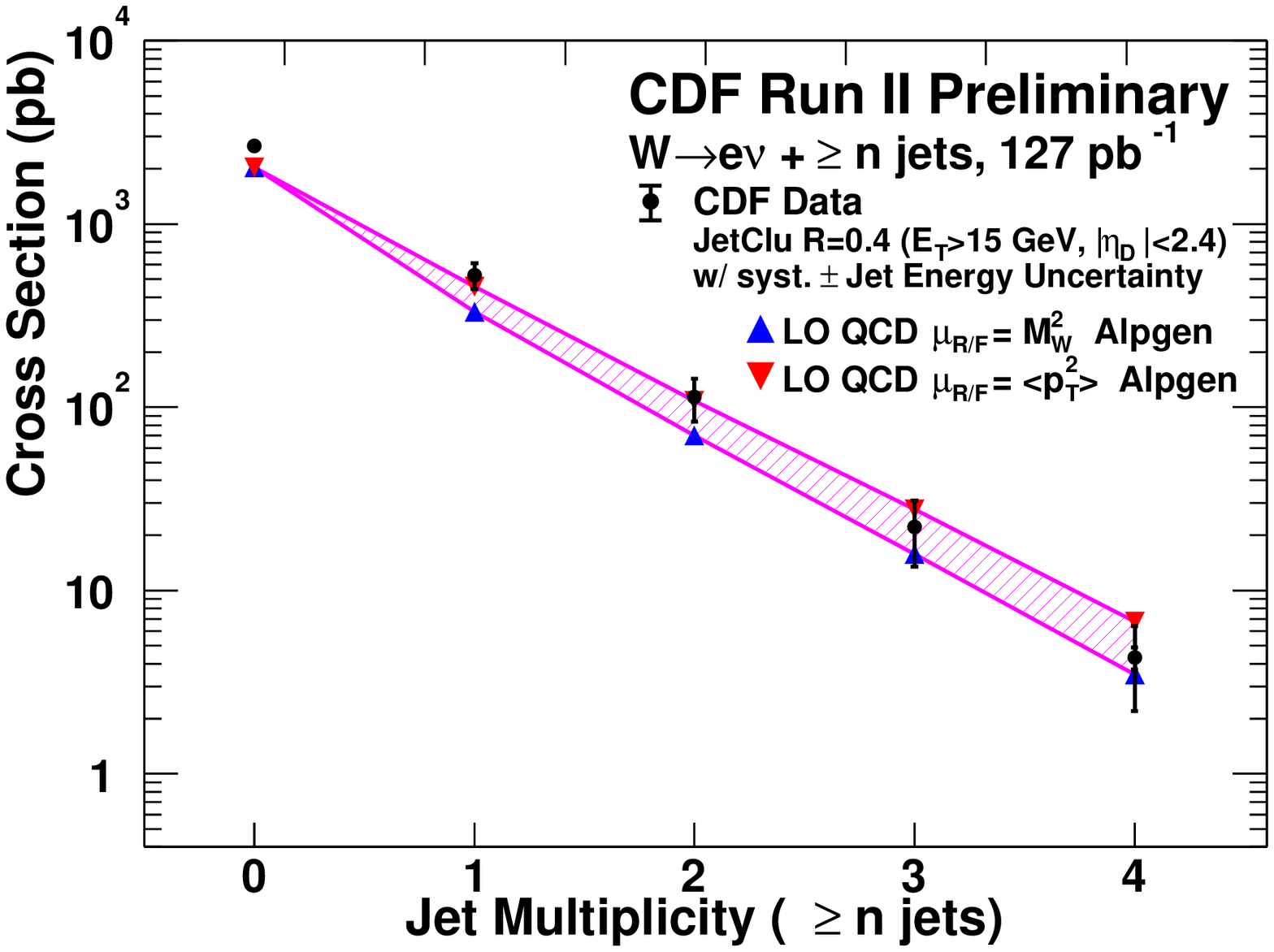}
\hskip -0.15in
\hfill
\includegraphics[scale=0.28]{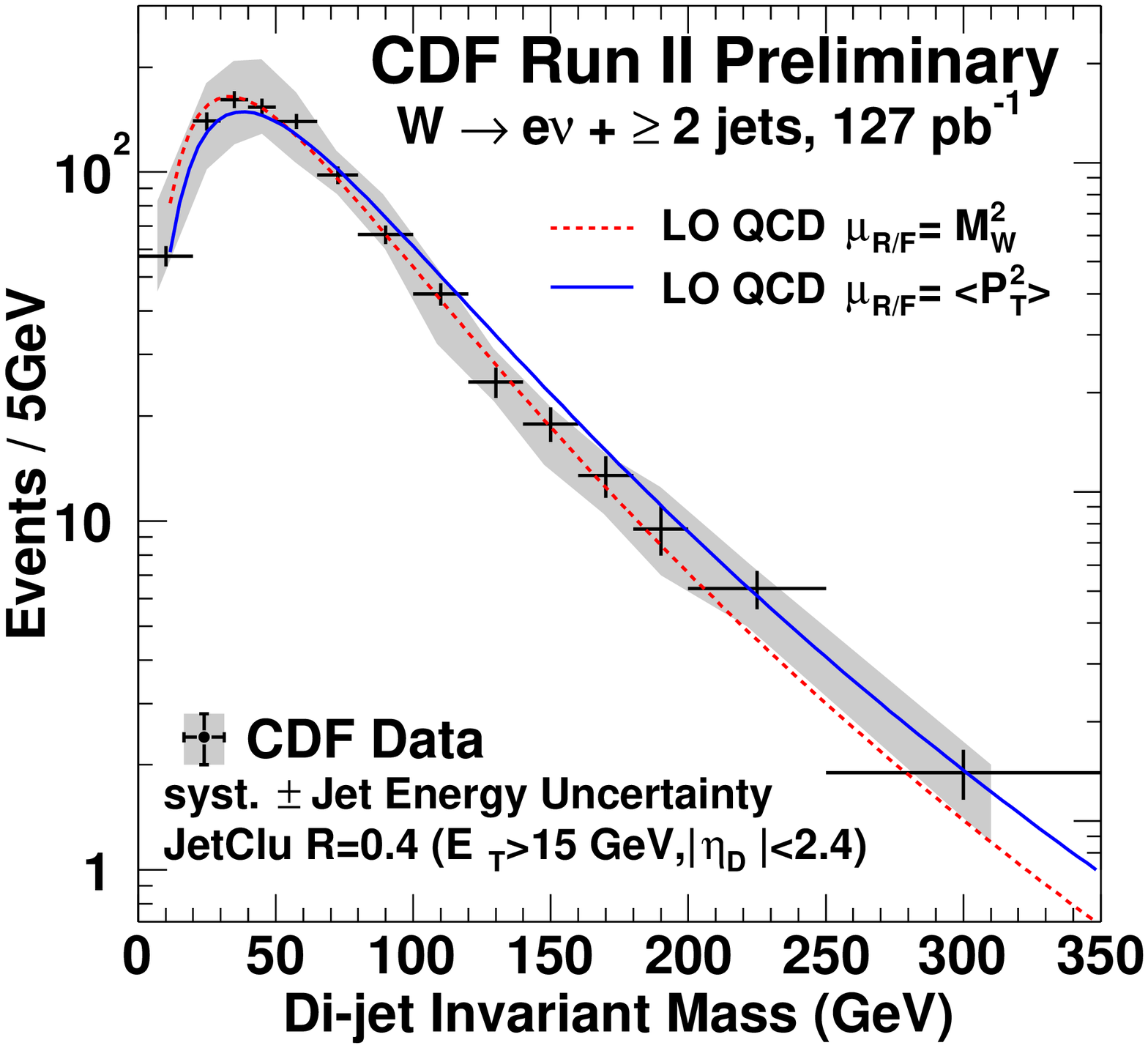}
}
\vskip -0.3in
\caption{\label{fig:data5} 
Left: CDF cross sections for $W$ + jets production vs. 
the inclusive number
of jets, compared to the LO prediction from {\sc alpgen+herwig},
for two choices of QCD scales.
Right: Invariant mass distribution of two leading jets
in CDF $W + \geq$2 jet events, compared to the same LO calculations.
Uncertainty from jet energy scale is shown by the band.}
\end{figure}

Production of electroweak bosons in association with several
jets has become an area of increasing interest for pQCD. 
The cross section for processes with $n$ jets is 
directly sensitive to $\alpha_s^n$, and the presence of
the heavy boson is expected to assure a more reliable
perturbative calculation. NLO predictions are available
up to $n=2$, while LO calculations ({\sc alpgen} \cite{alpgen}
code being primarily used by CDF and D\O) exist for larger
values of $n$. 
Frequently, such LO calculations are interfaced at the parton
level  to event generators such as {\sc pythia}
or {\sc herwig} \cite{herwig} to include the physics of soft emissions,
described through the parton-shower approach, and to facilitate
full simulation of experimental conditions following 
hadronization and modeling of underlying event.
Such ``enhanced-LO" procedures are of
obvious interest for investigations of many related 
physics processes, including top-quark and Higgs-boson
production, and searches for SUSY, for
which $W/Z$ + jets production represents a major background. 
Figure~\ref{fig:data5} shows CDF data on the $W + \geq n$ jets
cross section vs. the inclusive number of jets, and the dijet invariant
mass distribution in $W + \geq 2$ jet events, for jets with 
\Et\ $>$ 15 GeV and $|\eta|<2.4$, for a data sample of 127 \invpb.
The measurements are compared to expectations
from {\sc alpgen+herwig} (including
full detector simulation and event reconstruction) for
two choices of QCD scale of $M_W$ and 
$\sqrt{<p_{T,{\rm jet}}^2>}$;
the calculations bracket the data, but the theoretical
uncertainty is large at LO.

\begin{figure}
{
\includegraphics[scale=0.32]{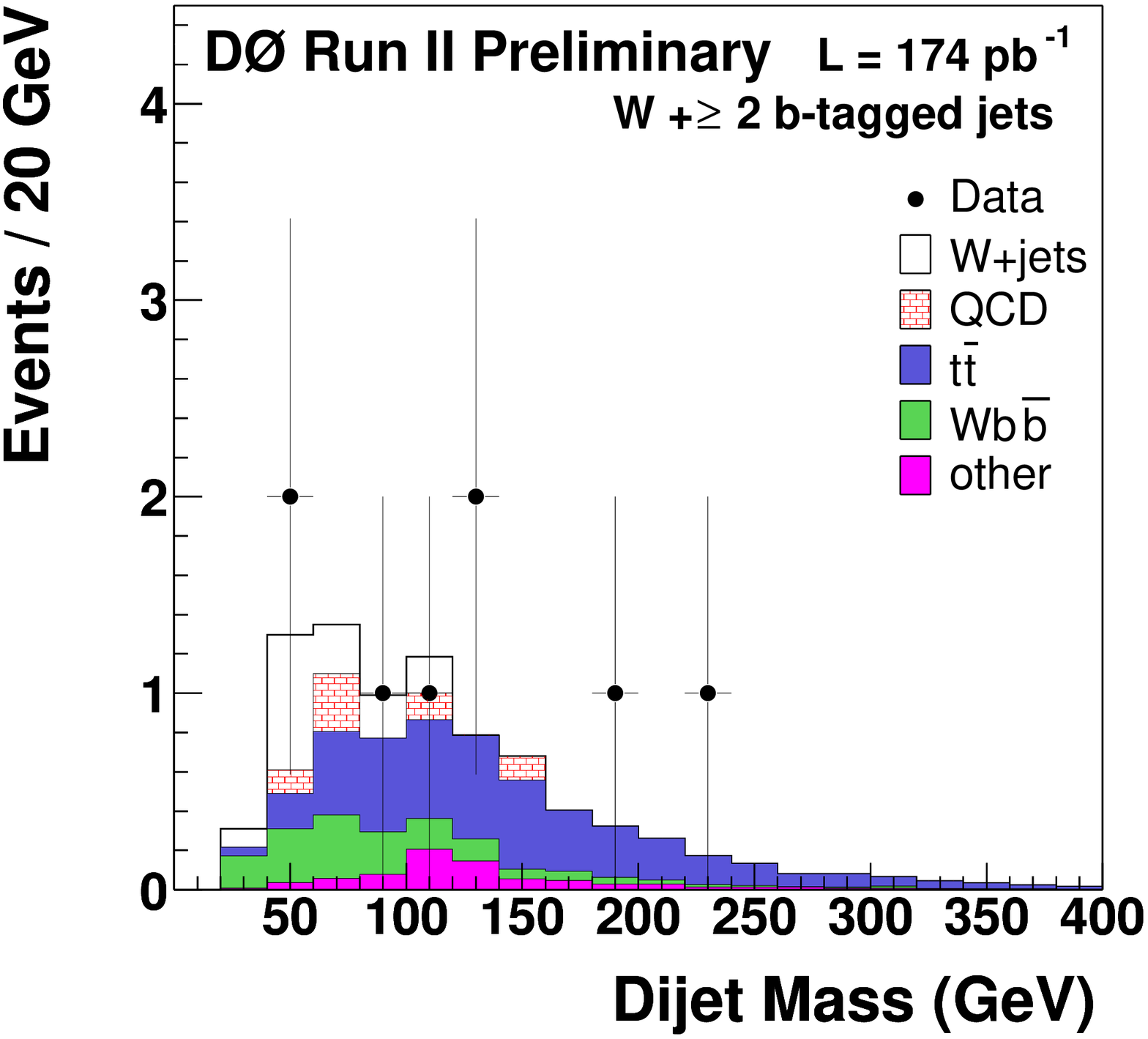}
\hskip 0.05in
\hfill
\includegraphics[scale=0.4]{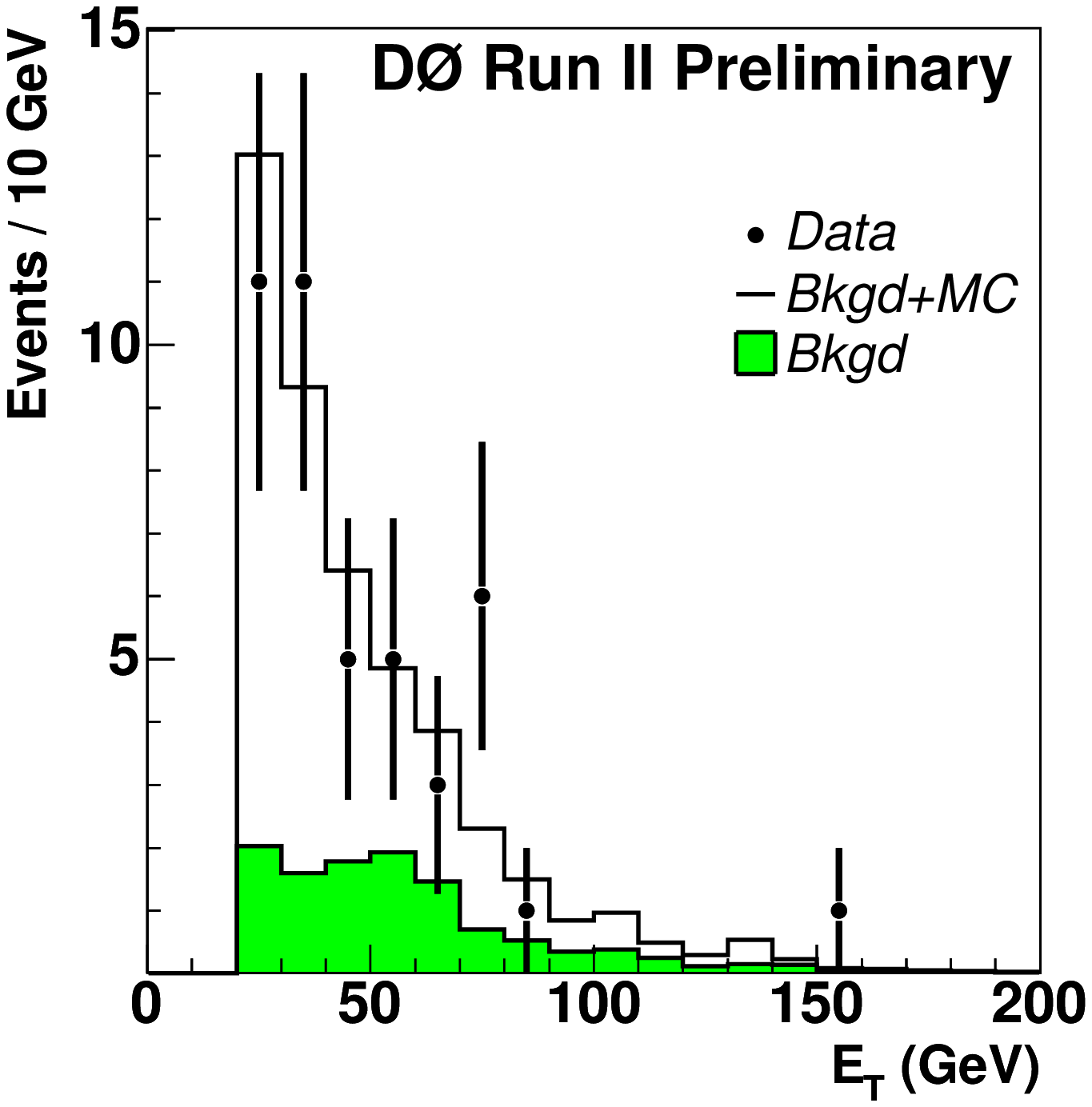}
}
\vskip -0.2in
\caption{\label{fig:data6} 
Left: Invariant mass distribution of two leading jets
in $W + \geq 2\,\, b$-tagged jet events measured by D\O. 
Data are compared to the simulated processes
using {\sc alpgen+pythia}.
Right: \Et\ spectrum of $b$-tagged jets associated with 
$Z$ production measured by D\O. The shaded histogram
represents background contributions estimated from data. 
The open histogram is the
sum of background and simulation of $Z + b$
production using
{\sc pythia}, normalized to data.
}
\end{figure}

CDF and D\O\ have developed algorithms to
identify jets originating from $b$-quarks ($b$-tagging).
Figure~\ref{fig:data6} illustrates results from
D\O\ for 174 \invpb\ of data on dijet mass distribution in $W + \geq 2 b$-tagged jet
events, compared to {\sc alpgen+pythia} predictions for several 
contributing processes, and for jet \pt\ distribution in 
$Z~+~b$ events compared to {\sc pythia} + experimental background. 
Good agreement is observed between data 
and expectations;
forthcoming high-luminosity data will challenge NLO calculations
with much reduced uncertainties.

\begin{figure}
\centerline{
\hskip -0.05in
\includegraphics[scale=0.7]{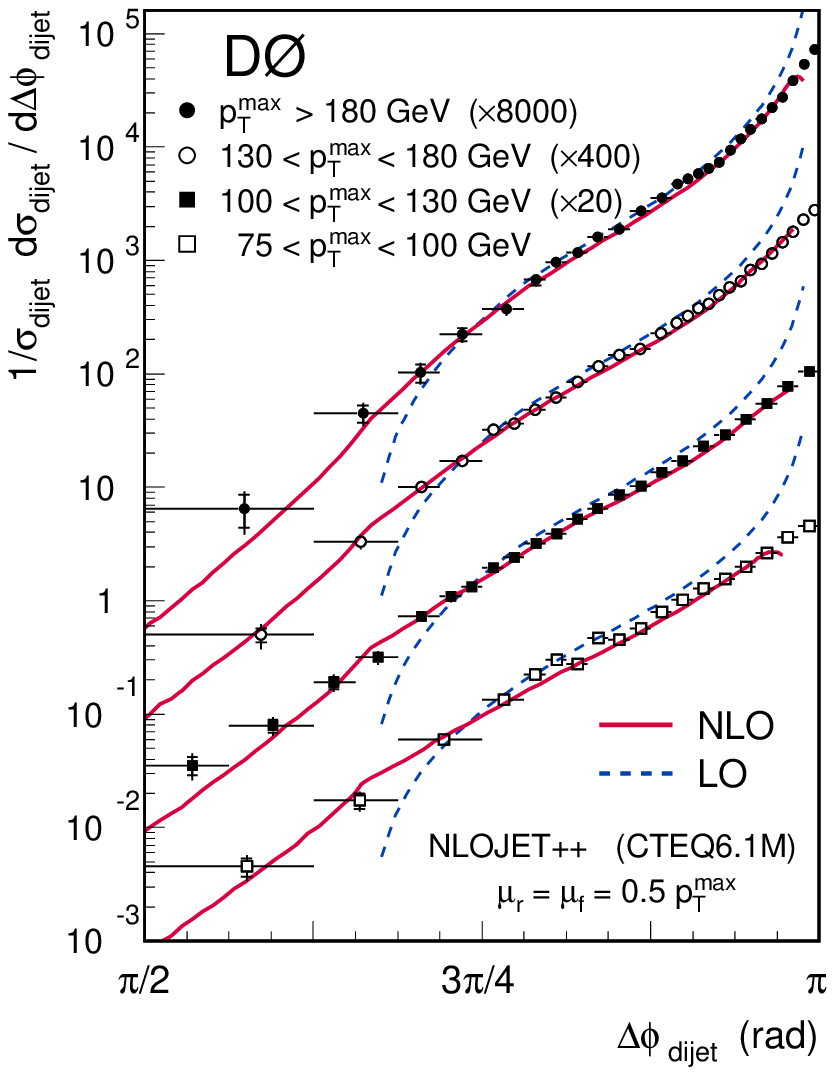}
\hskip 0.15in
\includegraphics[scale=0.7]{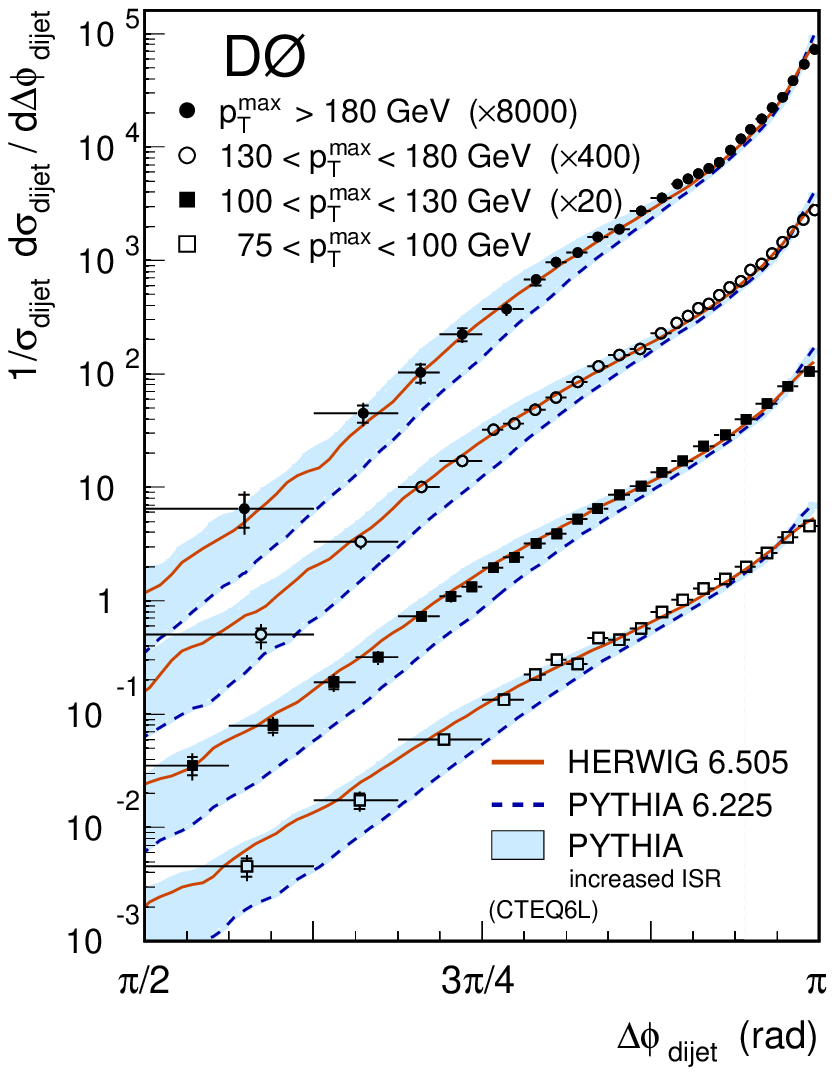}
}
\vskip -0.2in
\caption{\label{fig:data7} 
Left: \Dphi\ distributions from D\O\
in four regions of leading-jet \pt.
The solid (dashed) lines show the NLO (LO) pQCD predictions.
Right: Comparison of \Dphi\ distributions to results
from {\sc herwig} (solid line), default {\sc pythia}
(dashed line), and a possible variation in the level
of initial-state radiation in {\sc pythia} (band).}
\end{figure}

Aspects of radiation in hard-scatter events can be studied
without reconstructing multiple jets, by simply investigating
the azimuthal angle between the two leading jets, \Dphi. 
At lowest-order, the two jets are expected
to be produced back-to-back, \Dphi$\approx\pi$.
Thus, the distribution in \Dphi\  away from $\pi$ reflects 
the presence of higher-order radiation.
D\O\ data on \Dphi\  for 150 \invpb, normalized by the
dijet inclusive cross sections integrated over the 
same phase-space, are presented in
Fig.~\ref{fig:data7} for four ranges of \pt\ of the 
leading jet (the second leading jet was required to 
have \pt\ $>$ 40 \GeVc). Data is compared to pQCD
LO and NLO calculations
for 3-jet production using {\sc nlojet++} \cite{nlojet}. 
It is apparent
that the LO calculation has a very limited range of 
applicability, while the NLO prediction,
with up to 4 partons in the final state, 
provides a good
description over a large range of \Dphi.
{\sc pythia} and {\sc herwig} employ parton-shower
models to describe higher-order QCD effects.
{\sc herwig} agrees with
the data well over the entire range of \Dphi; 
the default version of
{\sc pythia} provides a rather poor representation (dashed
lines). However, the amount of radiation in 
{\sc pythia}-generated events can be adjusted, e.g.,
by varying the scale factor for the maximum \pt\ allowed 
in the initial-state parton shower ($\mathtt{PARP(67)}$).
The bands correspond to the variations in this parameter 
between the
default value of 1.0 and  4.0 (used in CDF Tune A). 
The higher values provide
a better description, and the data clearly can be used
to further tune the {\sc pythia} model.

\begin{figure}
\centerline{
\hskip 0.1in
\includegraphics[scale=0.35]{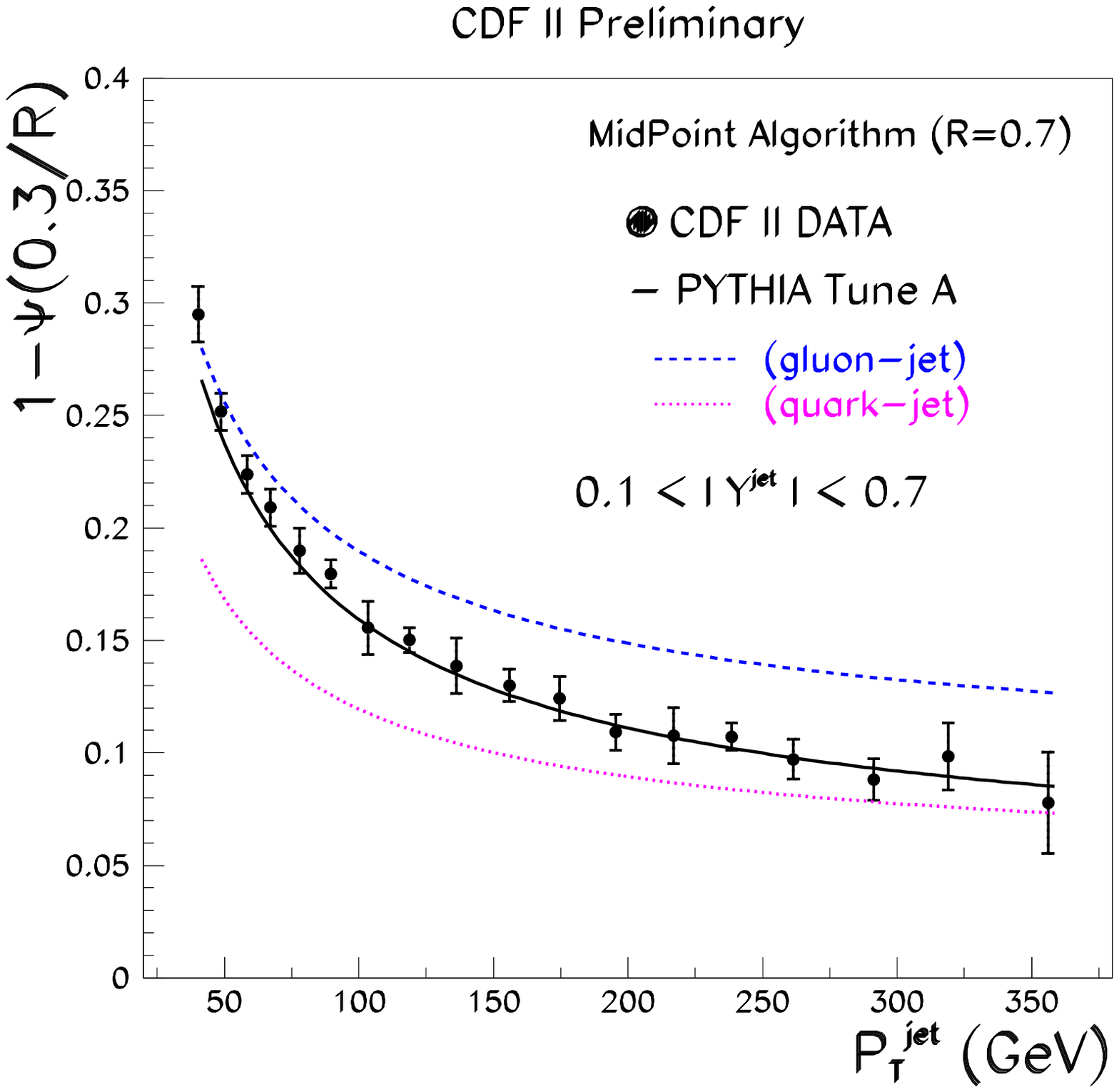}
\hskip -0.2in
\includegraphics[scale=0.35]{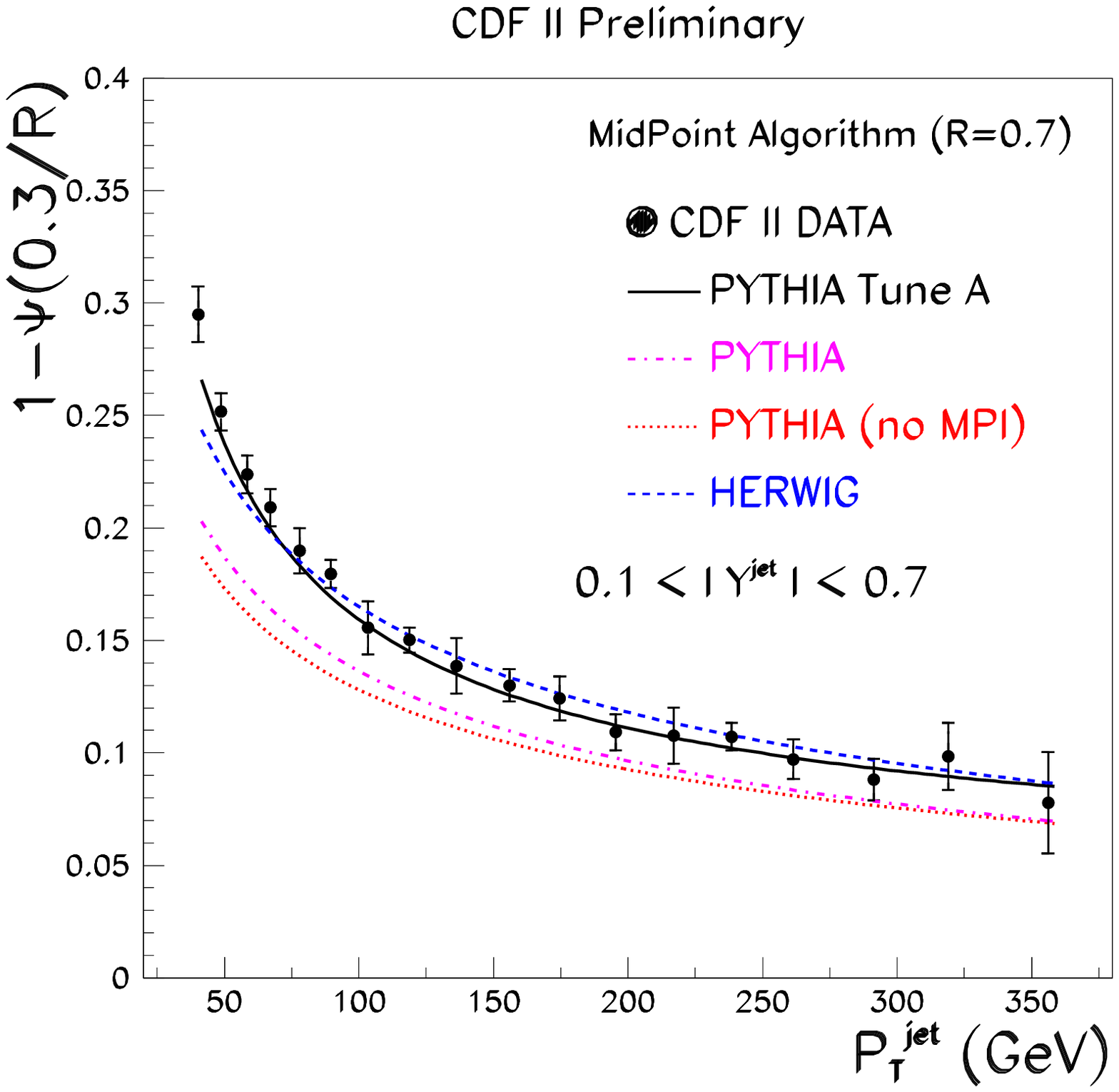}
}
\vskip -0.3in
\caption{\label{fig:data8} 
Left: CDF measurement of ``jet-\pt\ fraction'' in the
outer part of the jet cone, $0.3<r<0.7$, vs. jet \pt,
compared to prediction from {\sc pythia} (Tune A).
Expectations for pure quark and gluon-initiated jets
are also shown.
Right: Same data compared to results from
{\sc herwig}, and {\sc pythia} with default settings
and with no multiple-parton interactions.
}
\end{figure}

The distribution of energy flow within the jet cone
is sensitive to soft parton radiation.
Figure~\ref{fig:data8} shows CDF data on the fraction
of jet \pt\ integrated over the outer part of
the cone ($0.3<r<0.7$ for $R=0.7$ size cones) vs. jet \pt,
using the midpoint algorithm for 170 \invpb.
{\sc pythia}, with the Tune-A set of parameters, as well as
{\sc herwig}, properly describe the transition of jet
shapes from gluon-dominated at low \pt\ to quark-dominated
at large \pt; the default version of {\sc pythia}, and  
{\sc pythia} with multiple-parton interactions switched off,
undershoot the data.

\begin{figure}
\centerline{
\includegraphics[scale=0.45]{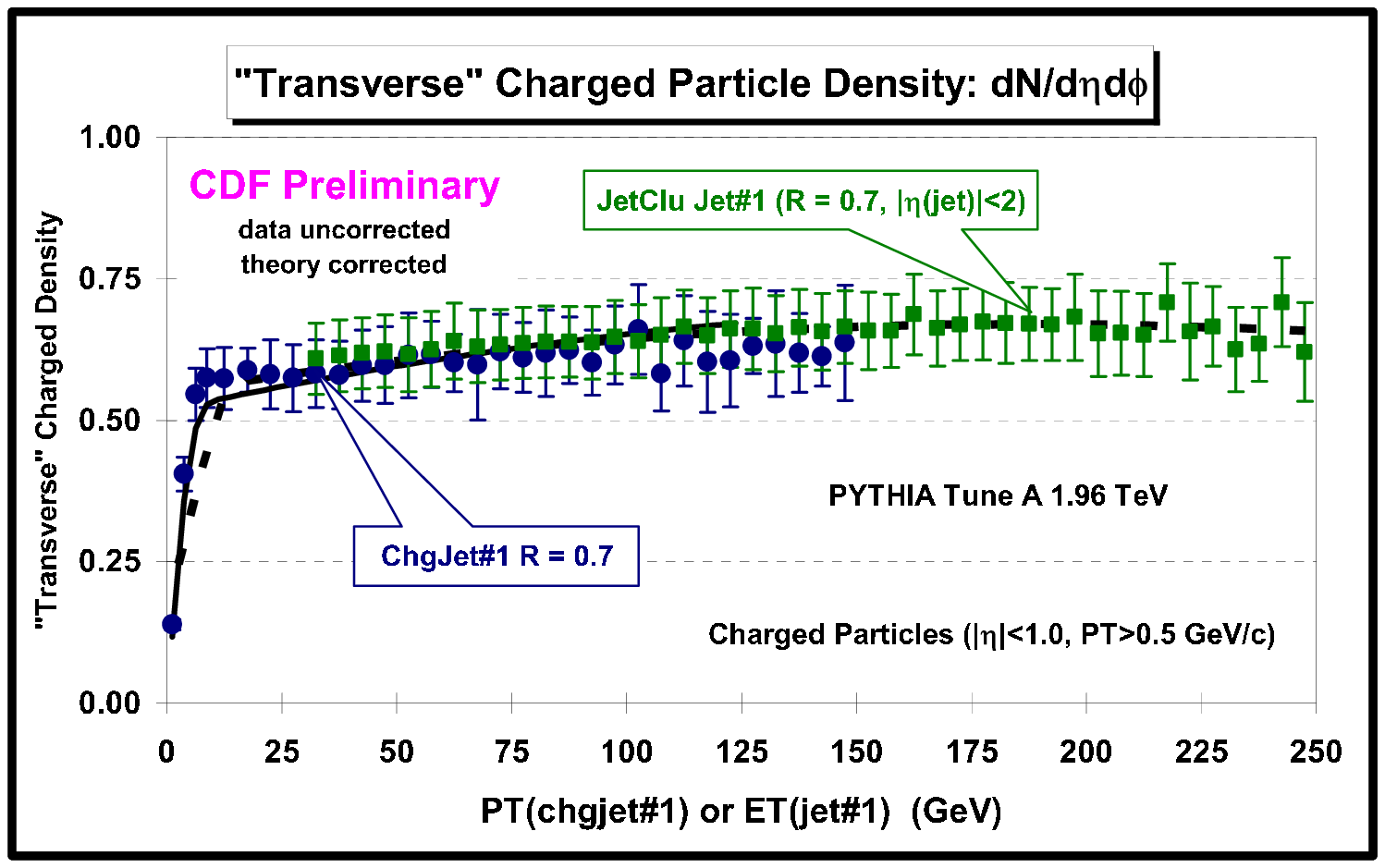}
\hskip 0.2in
\includegraphics[scale=0.3]{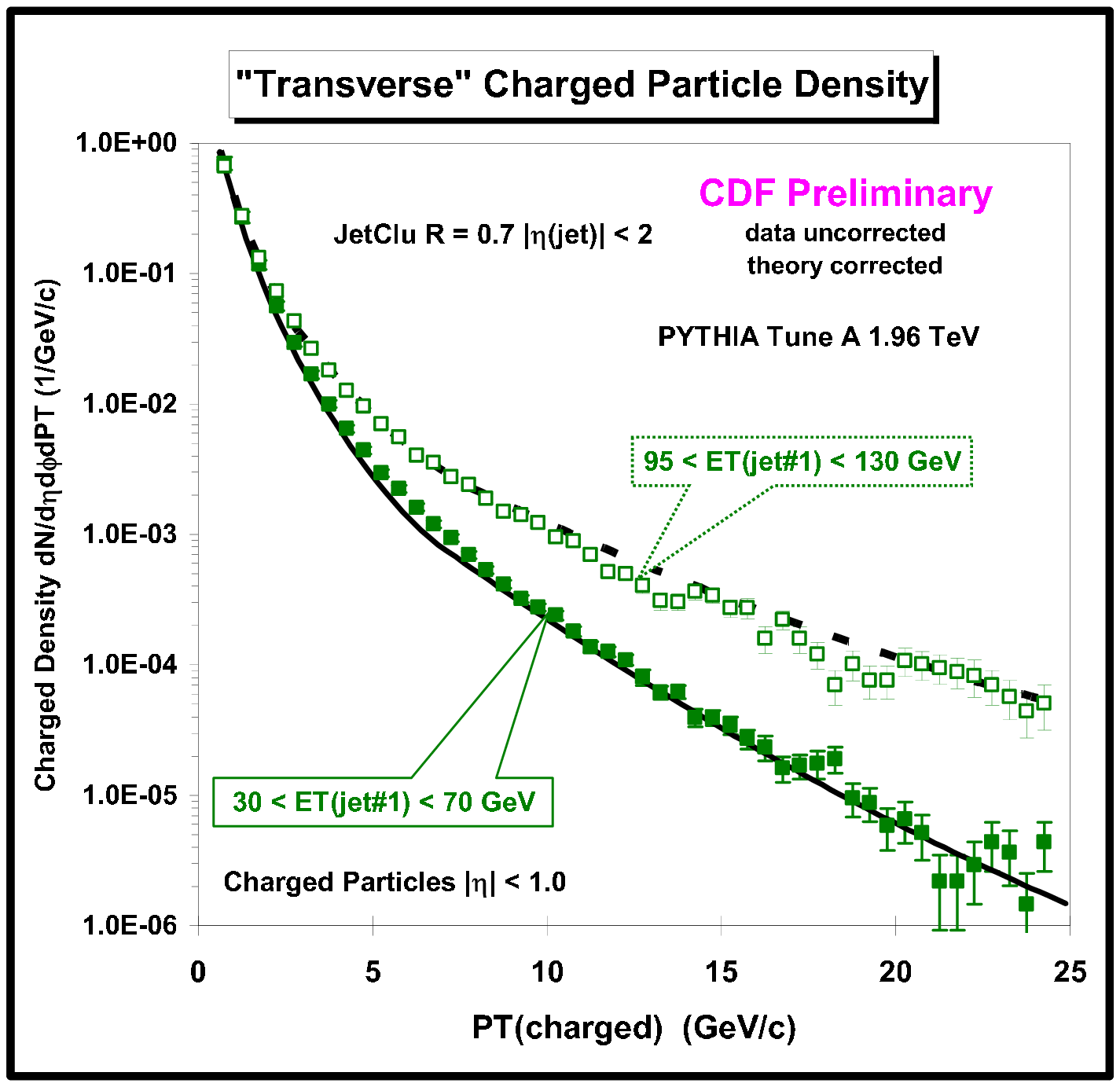}
}
\vskip -0.15in
\caption{\label{fig:data9} CDF studies of charged particle
density in the transverse region relative to the leading
jet (see text) vs. \pt\
of the jet (left), and vs. \pt\ of the charged
particles in the transverse region (right), 
compared to expectations from tuned 
{\sc pythia} (Tune A).}
\end{figure}

In addition to objects emerging from the hard scatter,
events contain soft particles produced through several
mechanisms, often collectively referred to as underlying
event (UE).
These include
fragments of beam remnants, multiple-parton
interactions, and initial and final-state radiation. 
Not all of these contributions are calculable within 
perturbation theory, and must therefore be modeled 
using Monte Carlo programs tuned to data.
Since particles from UE influence jet properties,
especially at low \pt, proper modeling of UE
is essential to account for such effects.
To enhance sensitivity to UE, in Run I
CDF studied \cite{r1cdfue}
distributions of charged particles with \pt\ $>$ 0.5 \GeVc\
and $|y|<1.0$ in the
region of the azimuthal
plane tranverse to the leading jet 
(defined using either charged particles or calorimeter information).
 Examples of similar Run II studies are presented in 
Fig.~\ref{fig:data9} for the density of charged particles
in the transverse region vs. \pt\ of the leading jet,
and vs. \pt\ of the charged particles in the transverse
region (in bins of \pt\ ot the leading jet).
The event generators require tuning of
phenomenological parameters to data to describe
the details of such distributions. 
CDF studies established a special tune of {\sc pythia}
parameters, Tune A, which provides 
a significantly improved description of UE properties
compared to default settings.
Inclusion of other processes in the tuning is
expected in the near future, and should help further constrain
the parameters and verify the universality of this
approach. 

A prediction from {\sc pythia} Tune A is
that, at $\sqrt{s}=14$ TeV, $\approx$12\% of all interactions
will result in hard scatters with \pt\ $> 10$ \GeVc\ ---
potentially an important effect on measurements in the 
high-luminosity environment of the LHC!

Further improvements in the
understanding of aspects of QCD at the Tevatron, and
development of related physics-analysis and simulation tools, will
undoubtedly be of great benefit to the physics program at the start-up 
of the LHC.

\bigskip

{\small It is a pleasure to thank my colleagues at CDF
and D\O\ for many helpful discussions.

\end{document}